\def\ii#1\ff{\textul{#1}}	

\def\diff{\mathrm{d}}

\def\A{\mathrm{A}}
\def\B{\mathrm{B}}
\def\C{\mathrm{C}}
\def\D{\mathrm{D}}

\def\vecp{\mathbf{p}}

\def\Ntest{N_{\textrm{test}}}

\def\bbsty#1#2#3{{\bf #1}, #2 (#3)}	
\def\textsty#1{\textit{#1}}	
\documentclass[prc,aps,twocolumn,psfig,preprintnumbers,amsmath,amssymb]{revtex4-1}
\usepackage{graphicx}\usepackage{epsfig}
\usepackage{amsmath}\usepackage{amssymb}\usepackage{amsfonts}
\usepackage{bm}
\usepackage{exscale}
\usepackage{dcolumn}    
\usepackage[dvipsnames]{xcolor}
\usepackage{soul}
\usepackage{ulem}
\usepackage{hyperref}
\begin{document}
%
%
\vspace*{2.5ex}
\title{Dynamic description of ternary and quaternary splits of \\heavy nuclear systems in the deep-inelastic regime}
%
%
\author{P. Napolitani$^1$, A. Sainte-Marie$^1$ and M. Colonna$^2$}
%
%
\affiliation{	$^1$ IPN, CNRS/IN2P3, Universit\'e Paris-Sud 11, Universit\'e Paris-Saclay, 91406 Orsay Cedex, France
		\\		$^2$ INFN-LNS, Laboratori Nazionali del Sud, 95123 Catania, Italy}
%
%
\begin{abstract}
	Colliding heavy nuclear systems in the deep-inelastic regime may undergo partitioning into multiple fragments when fusion can not be achieved.
	While multiple breakups are common at Fermi energy, they are rather exotic in the deep-inelastic regime, where density, excitation and, in general, transport conditions, are expected to be different.
	Abundant ternary and quaternary splits have been observed in recent experiments, for instance in symmetric semi-central and semi-peripheral collisions with heavy systems, like $^{197}\mathrm{Au}$ + $^{197}\mathrm{Au}$ at 15 MeV per nucleon.
	In these conditions, we undertook a microscopic description of the reaction dynamics.
	Relying on the full solution of the Boltzmann-Langevin equation implemented in the BLOB approach, we could follow in time the development of instabilities along deformation.
\\\vspace*{2.5ex}
%
\end{abstract} 
\maketitle
%
%
%
%
%
\section{Introduction \label{introduction}}

	Below the Fermi-energy domain,
around roughly $20$~MeV per nucleon, 
heavy-ion collisions 
are largely dominated by deep-inelastic processes~\cite{Volkov1978,Gobbi1980,Schroeder1984,Sanders1999}.
	In this framework, the development of large deformations, together with the presence of metastable conditions and mean-field fluctuations, has been related to a transition from fusion to binary processes and to a large variance of primary fragment properties~\cite{Umar2017}.
	However, when the system is too heavy to undergo fusion, alternative dynamical paths may explore a large variety of exotic deformed configurations even for non-peripheral impact parameters~\cite{Granier1988,Shvedov2010}.
	Stretched patterns may then rupture and be reabsorbed in binary configurations around the projectile-like (PLF) and the targetlike (TLF) fragment, or occasionally separate into multiple fragments.
	Along this transition from binary to multiple splits, deformed composite systems encounter unstable conditions where fluctuations and bifurcations of mean-field trajectories determine the fate of the process among a variety of possible exotic reseparation modes with many fragments.
	Multiple-split channels were already found in early experiments dedicated to heavy systems and largely discussed~\cite{Glassel1983,Vater1986,Qureshi1988,Casini1993,Charity1991,Stefanini1995,Blaich1996}.
	This body of data, completed by more recent exclusive experiments dedicated to very heavy nonfusing systems~\cite{SkwiraChalot2008,Cap2014}, revealed a fragmentation mechanism leading to fast ternary and quaternary splits~\cite{Wilczynski2010b} which are not mere isotropic emissions around the PLF and TLF, and which could be related to semi-central impact parameters~\cite{Wilczynski2010a} and very exotic noncompact shapes~\cite{Najman2015}.
	However, even though the exit channels with multiple fragments may suggest similarities between the deep-inelastic and the Fermi-energy domains, these regimes are related to rather different conditions.
	In fact, fragmentation processes at Fermi energy~\cite{EPJAtopicalWCI2006,EPJAtopicalNSE2014,Colonna2017} are mainly determined by mechanical (isoscalar) instabilities at low density~\cite{Colonna1998,Chomaz2004,Ducoin2007,Napolitani2017} and characterised by isospin drifts issued from a combination of density and isospin gradients~\cite{Baran2005,Li2008}, all this producing several fragments of comparable size in central collisions~\cite{Borderie2001,Borderie2018} or neck fragmentation~\cite{Montoya1994,Lionti2005,DiToro2006,Baran2012,  Hudan2012,DeFilippo2012,Brown2013,Jedele2017,Manso2017} in peripheral collisions.
	The same density and excitation conditions are unreachable at lower energy.
	For this reason, with respect to Fermi energy, ternary and quaternary channels in the deep-inelastic regime led to alternative interpretations which may be grouped in two options.
	A first description relies on a chiefly binary channel, where three or four fragments are achieved after a \textit{secondary} fission processes from two already well-separated heavier nuclei.
	Such process should be sufficiently fast to keep trace of the deformed shape imparted to the system by the collision, so that fission events from the PLF and TLF remnants are with larger probability aligned along the PLF-TLF separation axis.
	A second description advocates a \textit{direct} process involving large dynamical contributions so that, suddenly after the collision, one or two fragments form at midrapidity between the PLF and the TLF.
	Such mechanism, according to experimental results, should produce fragments of larger size in comparison to fragmentation processes at Fermi energies.
	The difference between the above two options is not necessarily the timing, which experimental analyses suggest to be very short under both interpretations, but the fact that the instability behind the process would act in two successive stages in the first case, or on the whole system at once in the second case.
	The situation may be even more complex, as these two pictures may coexist and smoothly alternate their contribution as a function of the impact parameter.
	This discussion calls for a thorough study of the involved mean-field dynamics, transport conditions, instabilities, shape fluctuations, and timing of the process from a microscopic point of view.

	In the following, we address this long-standing issue by carrying out a new theoretical analysis of the very heavy system $^{197}$Au$+^{197}$Au at 15~$A$MeV, which offers an extensively studied example~\cite{SkwiraChalot2008,Wilczynski2010b,Wilczynski2010a}. 
	Simulations based on stochastic mean field~\cite{Rizzo2014} as well as on molecular dynamics~\cite{Tian2010} have been already addressed to this system, leading to roughly comparable results when regarding the chronology of the dynamical process and impact-parameter dependencies.
	Rather, the difference depends upon the efficiency in combining fluctuations and collective effects in the dynamics. 
	The consequence is that, when fluctuations are approximated, like in Ref.~\cite{Rizzo2014}, the fragment partitioning process appears inhibited by the resilience of the mean-field with decreasing incident energy.
	Eventually, at low energy, fragment formation can be definitely compromised if the treatment of fluctuation growth is incomplete.
	Molecular-dynamics approaches show the opposite effect: fluctuations naturally arise from nucleon-nucleon correlations, but mean-field resilience is underestimated with the consequence of approximating collective effects on the density distributions.
	In particular, even though the time scales may look similar in some respect, the deformation of composite systems leading to multi-fragment configuration are not comparable to mean-field simulations and present a different evolution with impact parameter.
	Very elongated configurations are achieved in mean-field approaches~\cite{Rizzo2014}, while molecular dynamics calculations like in Ref.~\cite{Tian2010} seem to favour more compact or fractal-like configurations. 

	In this work, in the same spirit of Ref.~\cite{Rizzo2014}, where the $^{197}$Au$+^{197}$Au system was also simulated, we employed a stochastic one-body approach which, in this new attempt, adopts a full treatment of the Boltzmann-Langevin equation in three dimensions, condensed in the BLOB model~\cite{Napolitani2013,Napolitani2017}.
	Such approach handles shape fluctuations in the dynamical evolution more consistently, while keeping an efficient description of collective behaviour relying on the mean-field formalism.
	As a result, it describes the separation of the system into more than two fragments and allows to investigate the nature of instability which determines the breakup of the system in numerical sinulations, as well as from a linear-response-theory point of view.
	We show in the following that, according to this approach, deformed patterns produced in very heavy nonfusing systems may range from annular shapes to neck threads between the reaction participants as a function of the impact parameter and result into three or four fragments.
	We also determine that the breakup process progresses from surface instabilities of Rayleigh type near nuclear saturation density with a growth time which suggests an overlap, or a gradual transition, between a direct process and a secondary nonisotropic fission process.
	Such phenomenology should characterise heavy systems in the deep-inelastic regime in general, where instabilities arise from highly deformed configurations and not, like at Fermi energy, from reaching densities far below saturation.
	We conclude our analysis by defining dispersion relations for the multi-fragment process which allows to clearly distinguish the deep-inelastic regime from the Fermi-energy domain.

\section{Boltzmann-Langevin modelling of instabilities below Fermi energy  \label{model}} 

	To simulate stretched patterns in nuclear systems and the corresponding instabilities which trigger fragmentation and clustering, we followed two steps.

	First, we focused on nuclear matter, where clusterisation is a general catastrophic process characterising Fermi liquids~\cite{Pines1966} as a response to fluctuations.
	Fluctuations involve either neutrons and protons moving in phase (isoscalar) or out of phase (isovector). 
	clusterisation can be described as progressing from zero-sound propagation, and resulting into ripples in the density landscape. 
	In this case, the corresponding isovector and isoscalar amplitudes should evolve according to the symmetry energy and the dispersion relation for unstable isoscalar modes, respectively.
	The model has been tested to ensure that fluctuation amplitudes are consistent with analytic expectations from Fermi liquids~\cite{Napolitani2017}.
	In nuclear matter, the isoscalar mechanism yielding clusterisation requires low-density conditions and negative compressibility of matter resulting into volume instabilities of spinodal type, or it is entertained by other close regions of the equation of state, like Landau-damping regimes~\cite{Chomaz2004}.

	In a second step, the model prepared to describe nuclear-matter conditions was applied to rapidly-evolving self-bound finite systems, as those produced in dissipative heavy-ion collisions. 
	The microscopic dynamical approach is naturally suited for handling out-of-equilibrium conditions. 
	At variance with nuclear matter, in this application to dissipative heavy-ion collisions where necks and deformed topologies may form, also surfaces should be affected by instabilities and contribute to clusterisation.
	In particular, the dynamics of a nuclear neck is characterised by the Plateau-Rayleigh instability~\cite{Rayleigh1882}, like in the classical picture of fission~\cite{Vandenbosch1973,Brosa1990}.

	A mean-field approach (MF) is well suited for satisfying the above requirements: the zero-sound propagation, the description of nuclear-interaction properties like the symmetry energy and linear-response behaviour, the description of collective motion, and the handling of surface instabilities.
	To additionally describe clusterisation, further extensions beyond mean-field should be introduced to handle large-amplitude dynamics and address highly non-linear regimes.
	We employed a stochastic treatment to recover upper orders beyond two-body correlations in an approximated form (or, in terms of BBGKY hierarchy for the density matrices~\cite{Balescu1976}, we had to introduce correlations beyond the kinetic-equation level).
	A way to achieve this goal is to introduce nucleon-nucleon (N-N) correlations in the implementation of the N-N collision integral and exploit them by handling a stochastic ensemble of several MF trajectories.
	This treatment introduces a stochastic source $\delta I_{\textrm{coll}}^{(n)}$ of vanishing mean around the average collision contribution $\bar{I}_{\textrm{coll}}^{(n)}$ 
which can generate fluctuations intermittently in time.
	As a consequence, a single MF trajectory $\rho_1$ taken at a given time, separates into a subensemble  $\{ \rho_1^{(n)};\, n=1, \dots, \textrm{subens.} \}$ of new trajectories at a successive time, leading to a scheme which is close to stochastic TDHF~\cite{Ayik1988,Lacombe2016}:
\begin{equation}
	i\hbar\frac{\partial\rho_1^{(n)}}{\partial t} \approx [k_1^{(n)}+V_1^{(n)} , \rho_1^{(n)}] 
		+ \bar{I}_{\textrm{coll}}^{(n)} + \delta I_{\textrm{coll}}^{(n)}   \;.
\label{eq:STDHF}
\end{equation}
	The Wigner transform of Eq.~(\ref{eq:STDHF}) yields the Boltzmann-Langevin (BL) equation~\cite{Reinhard1992}, in terms of an effective Hamiltonian $h^{(n)}$ acting on a corresponding ensemble of distribution functions $f^{n}$.
	To solve the BL equation in full one-body phase space, we employ the Boltzmann-Langevin one body treatment (BLOB), resulting in the following set of BLOB equations~\cite{Napolitani2013,Napolitani2017}:
\begin{eqnarray}
	&&\frac{\partial f^{(n)}}{\partial t} - \{h^{(n)} , f^{(n)}\} 
		= I_{\textrm{UU}}^{(n)} + \delta I_{\textrm{UU}}^{(n)} = 
\notag\\
	&&= g\int\frac{\diff\vecp_b}{h^3}\,
	\int
	W({\scriptstyle\A\B\leftrightarrow\C\D})\;
	F({\scriptstyle\A\B\rightarrow\C\D})\;
	\diff\Omega
\;,
\label{eq:BLOB}
\end{eqnarray}
where $g$ is the degeneracy factor, $W$ is the transition rate in terms of relative velocity between the two colliding phase-space portions, and $F$ handles the Pauli blocking of initial and final states over their full one-body phase-space extensions.
	The distribution functions $f^{n}$ correspond to Fermi statistics at equilibrium and replace the Slater determinants in Eq.~(\ref{eq:STDHF}). 	
	The residual contributions in Eq.~(\ref{eq:STDHF}) are replaced by modified Uehling-Uhlenbeck (UU) terms, where each single in-medium collision event acts on extended equal-isospin phase-space portions, large enough so that the occupancy variance in $h^3$ cells corresponds to the scattering of two nucleons; this variance should equal $f(1-f)$, in order to strictly avoid any violation of Pauli blocking at each single scattering event~\cite{Rizzo2008}.

	The approach of Eq.~(\ref{eq:BLOB}) can be applied to investigate the leading instability and the role of dissipation in very deformed nuclear systems formed in heavy-ion collisions.
	Especially below the Fermi-energy domain, it provides a well-suited framework to study the interplay between the propagation of fluctuations and mean-field resilience, which introduces a smoothing contributions to fluctuations. 

	For comparison, a simplified approach where the fluctuations induced by the collision term of Eq.~(\ref{eq:BLOB}) are projected and implemented only in coordinate space is provided by the SMF~\cite{Colonna1998} model, as adopted in Ref.~\cite{Shvedov2010}.
	We refer to SMF calculations in the following. 

\section{Selection of timing and the collision configuration  \label{collision}} 
%
%
\begin{figure}[b!]\begin{center}
    \includegraphics[angle=0, width=.85\columnwidth]{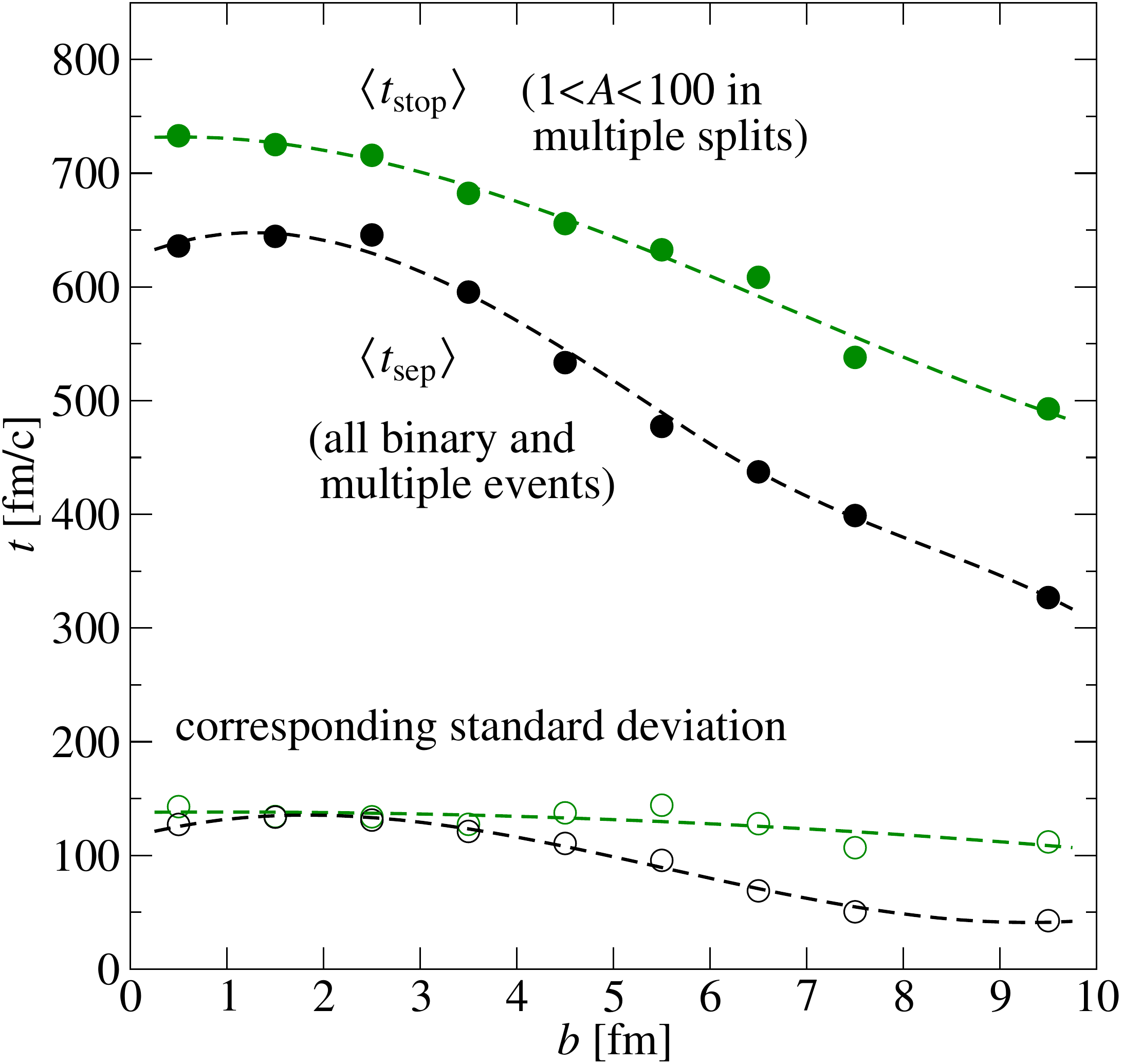}
\end{center}\caption
{
	Fragment chronology as a function of the impact parameter $b$ for $^{197}$Au$+^{197}$Au at 15~$A$~MeV calculated with BLOB.
	Average PLF-TLF re-separation time $\langle t_{\textrm{sep}}\rangle$ obtained considering all events 
and average emission time of the last fragment $\langle t_{\textrm{stop}}\rangle$ (with $1<A<100$) for events with fragment multiplicity $M\ge 3$ (filled symbols). 
	Open symbols of corresponding colour indicate the associated standard deviation.
	Time is measured starting from $t_0$. 
	Lines are fits.
}
\label{fig_time_split_vs_b}
\end{figure}

	We focus on the system $^{197}$Au$+^{197}$Au at 15~$A$MeV, adopting the BLOB approach.
	In the present study we adopt the same mean-field description as in Ref.~\cite{Rizzo2014}, used in the SMF model~\cite{Colonna1998} to simulate the same reaction. It is based on a simplified SKM* effective interaction~\cite{Guarnera1996, Baran2005} with incompressibility modulus $k=200$~MeV and a linear parameterisations for the surface symmetry energy.
	The mean field is sampled using $\Ntest =40$ test particles per nucleon. 
	A screened in-medium N-N cross section (from Ref.~\cite{Danielewicz2011}) is implemented in the transition rate $W$.
	The BLOB calculation is stochastic and requires therefore several events.
	500 BLOB events have been ran for successive impact parameter intervals $b=[0,1],[1,2], ... ,[7,8]$~fm; 700 events were ran for the interval $b=[9,10]$~fm and a larger statistics of 4000 events was chosen for the interval $b=[5,6]$~fm (as discussed in the following, this interval is associated to a large rate of quaternary splits and requires therefore special attention).

	We take the intervals $b=[5,6]$ and $b=[9,10]$~fm as representative for `semi-central' and `semi-peripheral' collisions, respectively.
	Similarly to the simulation strategy used in Ref.~\cite{Napolitani2015}, the collision process is sampled from the initial time $t_0$ until the so-called stopping time $t_{\textrm{stop}}$.
	$t_0$ corresponds to a displacement of 18~fm measured along the beam axis between the two gold nuclei; the time $t$ is calculated starting from $t_0$.
	$t_{\textrm{stop}}$ is the time corresponding to the separation of the last fragment in the dynamical process.
	Technically, because $t_{\textrm{stop}}$ can not be known a priori, it is searched in a window ranging from 400 to 1000~fm/c: the BLOB calculation runs until 1000~fm/c, and it is then rewound back to $t_{\textrm{stop}}$, which is therefore set differently for each single event.
	For the intervals $b=[6,7],[7,8]$ and $b=[9,10]$~fm the $t_{\textrm{stop}}$ window range is defined as $[300,900],[200,800]$ and $[200,800]$~fm/c, respectively.

%
%
\begin{figure}[t!]\begin{center}
    \includegraphics[angle=0, width=.9\columnwidth]{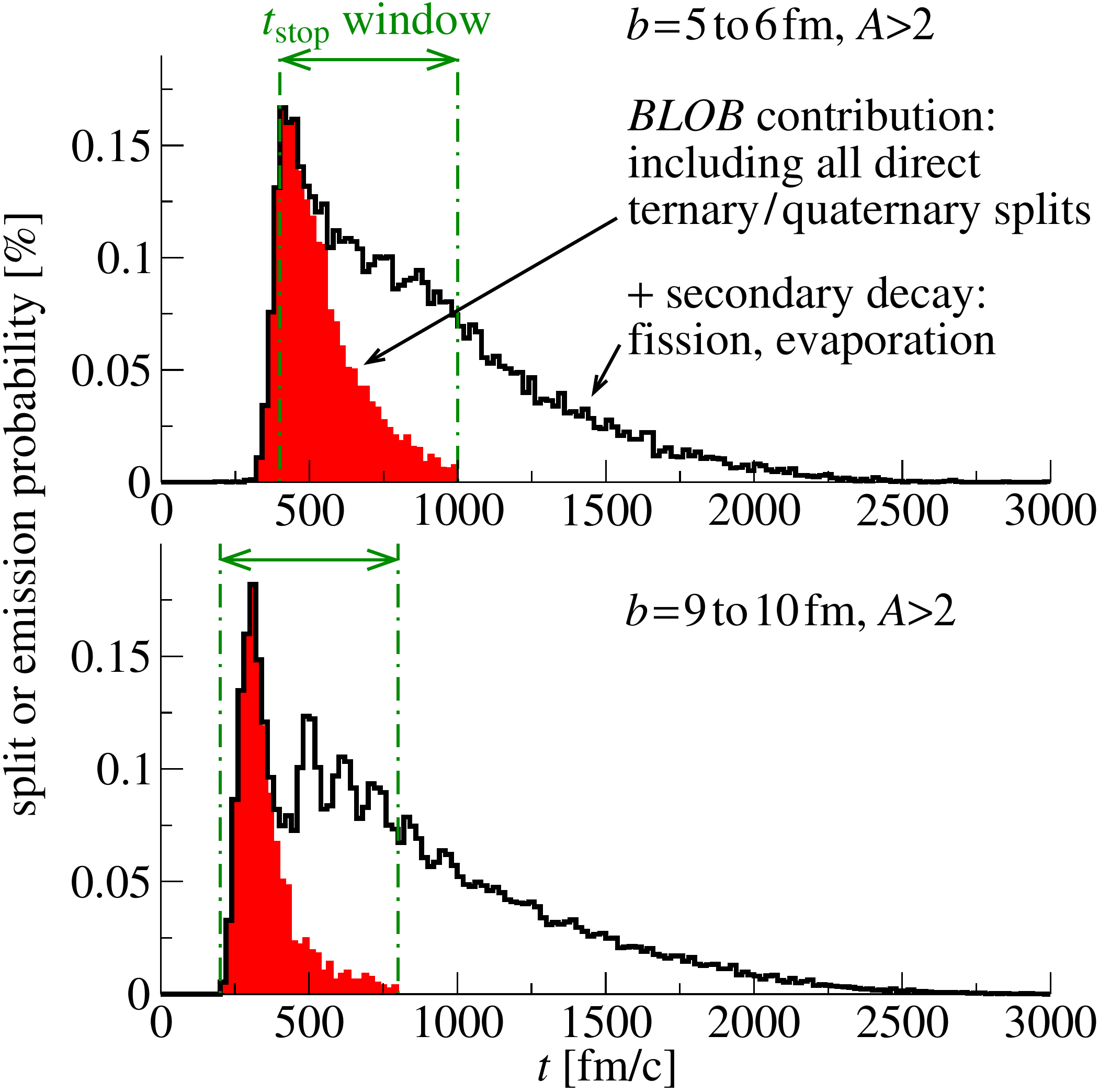}
\end{center}\caption
{
	Probability of split or emission of fragments with $A>2$ as a function of time for $b=[5,6]$~fm (top) and $b=[9,10]$~fm (bottom).
	The BLOB contribution up to $t_{\textrm{stop}}$ (where direct ternary and quaternary splits take place) is the red filled area, while the full distribution accounts for additional secondary decays.
}
\label{fig_Probability_split_b5to6_b9to10}
\end{figure}
%
%
\begin{figure*}[t!]\begin{center}
    \includegraphics[angle=0, width=.74\textwidth]{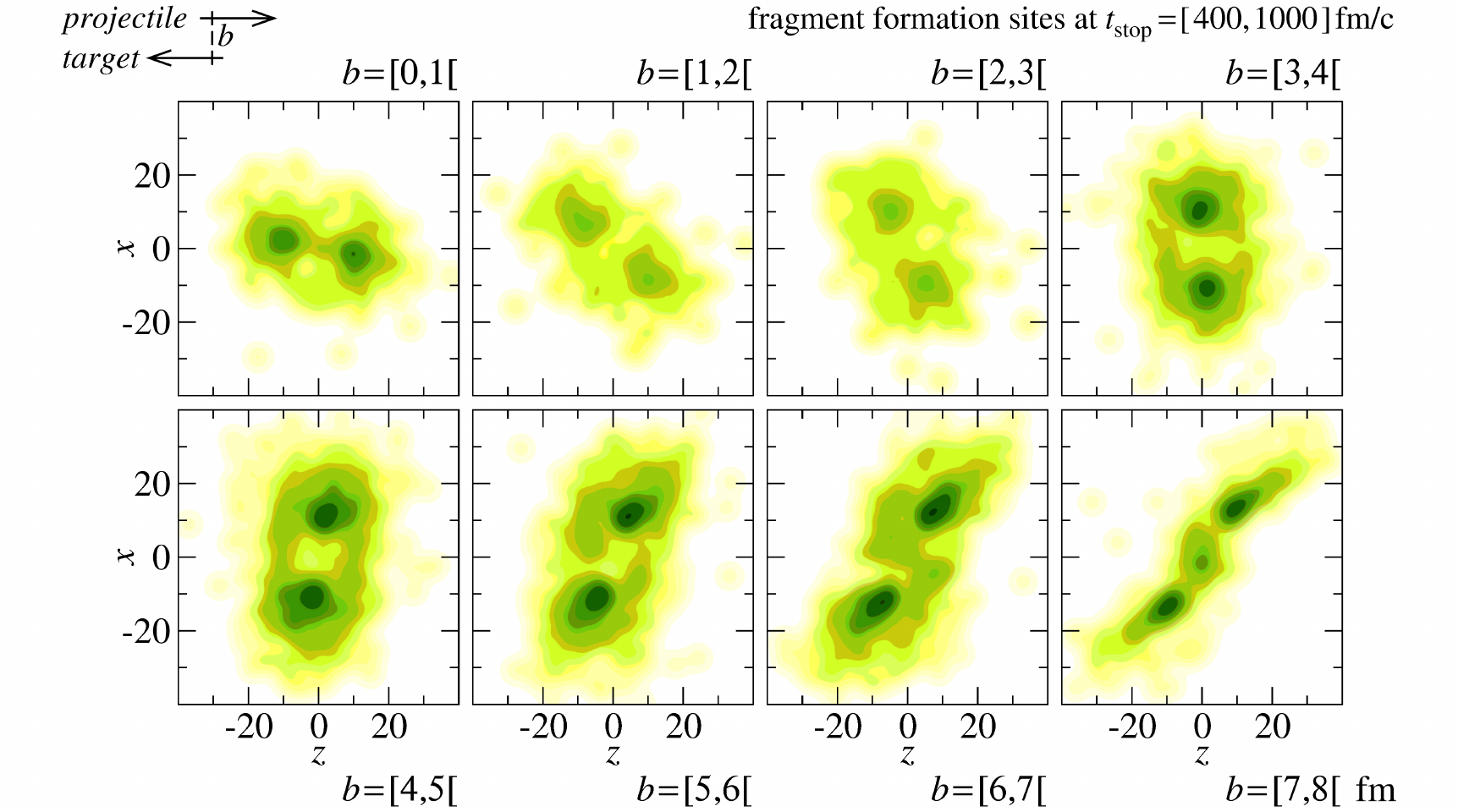}
\end{center}\caption
{
	Projection on the reaction plane of the distribution of fragment formation sites at $t_{\textrm{stop}}$ (not to be confused with a density distribution), for events corresponding to different $b$ ranges.
}
\label{fig_fragment_formation_sites_vs_b}
\end{figure*}

	The BLOB approach succeeds in describing mechanisms leading to multiple splits, which were not achieved in the previous SMF simulations~\cite{Rizzo2014}.
	Fig.~\ref{fig_time_split_vs_b} indicates that the average emission time of the last fragment $t_{\textrm{stop}}$ occurs right after the re-separation time $t_{\textrm{sep}}$ of the PLF and the TLF.
	Both times $t_{\textrm{stop}}$ and $t_{\textrm{sep}}$ decrease with increasing impact parameter $b$ and lose this correlation at small $b$ values.
	The gradual loss of sensitivity to $b$ with increasing centrality is expected at low incident energy for increasingly dissipative conditions~\cite{Napolitani2010}.

	Fig.~\ref{fig_Probability_split_b5to6_b9to10} studies the probability of split as a function of time for semi-peripheral collisions in the interval $b=[5,6]$ and $b=[9,10]$~fm.
	The red filled areas represents the contribution calculated within BLOB where all direct ternary and quaternary splits take place; these splits are all registered in the selected $t_{\textrm{stop}}$ window.
	Each BLOB event follows a different mean-field trajectory, resulting in an ensemble of possible close representations of the exit channel for one given initial state.
	These representations are selected at $t_{\textrm{stop}}$ and used to carry on the calculation with the decay afterburner Simon~\cite{Durand1992}, which propagates in time all fragments and particles in their mutual Coulomb field while processing secondary-decay events in flight.
	This procedure ensures that the action of the Coulomb field is correctly reflected in the kinematics and allows to associate times also to the secondary-decay products.
	The black histograms in Fig.~\ref{fig_Probability_split_b5to6_b9to10} illustrates the full distribution of fragment and cluster emission probability resulting from the complete calculation (i.e. BLOB continued with the secondary decay).
	Recalling the scenario depicted in the introduction, we may distinguish between {\it direct} (or {\it dynamical}) splits, when referring to early fragment formation described within the BLOB calculation, and {\it secondary} decays, when referring to later emissions described within the Simon afterburner. 
	Already at this stage, we may point out that there is no evidence of two separate regimes in the split rate distribution shown in Fig.~\ref{fig_Probability_split_b5to6_b9to10}.
	The dynamical and secondary-decay regimes are in fact overlapping in time.
	As discussed in the following, what distinguishes them is the type of mechanism and the associated kinematic properties.
	The study or Ref.~\cite{Tian2010} restricted to the dynamical contribution alone is quite comparable to the present BLOB calculation when regarding fragment separation times for non-central collisions, as well as for the primary-fragment emission rate if secondary decay is not accounted for.

	The distinction between direct and secondary mechanisms, introduced above, is essentially based on the reaction chronology which, as mentioned, does not imply distinct time regions when averaging over all events. 
	Its association to a distinction between dynamical and statistical mechanisms comes naturally from the fading of dynamical effects in time, but in the present two-step simulation strategy is also imposed by construction.
	In reality, while the direct process progresses immediately from the deformation imposed by the collision to the entire system, the secondary process may also inherit deformations affecting the PLF and TLF, when those latter are too heavy to relax in shape.
	However, within the two-step approximation, the decay of the PLF and TLF is described without taking into account any possible deformation when processed outside of the BLOB $t_{\textrm{stop}}$ window through the secondary-decay afterburner, with the consequence of losing possible long-lasting memory effects on the kinematics.

\section{Fragment observables} \label{results}

	We review thereafter kinematic and production observables calculated for the system $^{197}$Au$+^{197}$Au at 15~$A$~MeV.
	In particular, we focus on semi-central collisions where direct quaternary splits have such a large rate to compete with ternary splits, and dedicate some attention to semi-peripheral collisions, where the fading quaternary mechanism and abundant ternary channels arise from neck dynamics.

%
%
\begin{figure}[b!]
\begin{center}
    \includegraphics[angle=0, width=.9\columnwidth]{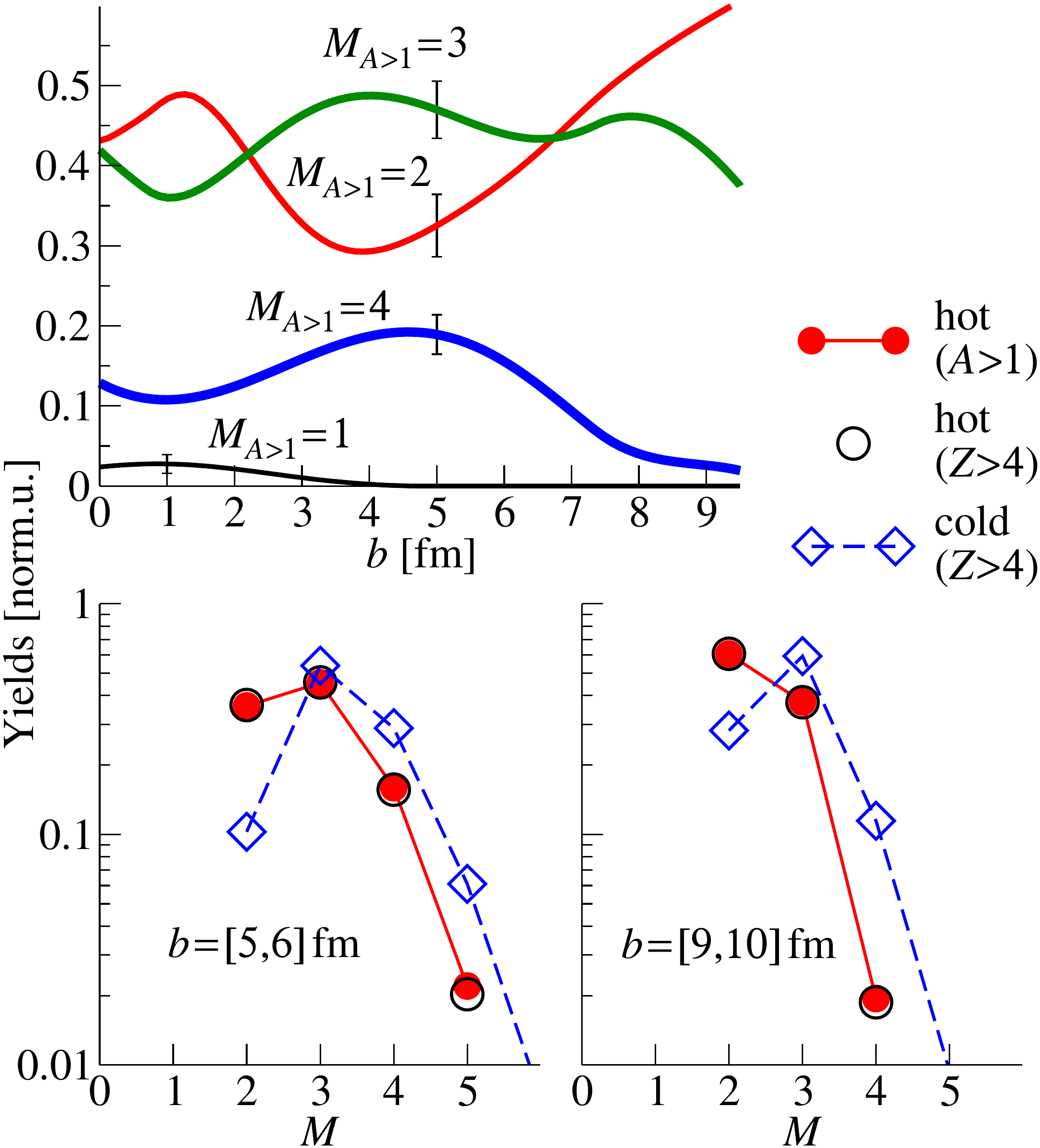}
\end{center}\caption
{
Top. Fragment-multiplicity contributions (including all $A\!>\!1$) represented by fitted average fragment yields as a function of $b$, calculated with BLOB up to the time $t_{\textrm{stop}}$ and normalised to the number of events.
Error bars indicate uncertainties from fits and calculation statistics.
Bottom: selecting the intervals $b=[5,6]$~fm (left) and $b=[9,10]$~fm (right), fragment-multiplicity distributions (selecting all $A\!>\!1$) at $t_{\textrm{stop}}$ normalised to the number of events processed with BLOB, and distributions (selecting all $Z\!>\!4$) for cold fragments after secondary-decay. 
}
\label{fig_M_b_stat_survey}
\end{figure}
%
%
%
\begin{figure*}[]
\begin{center}
    \includegraphics[angle=0, width=.95\textwidth]{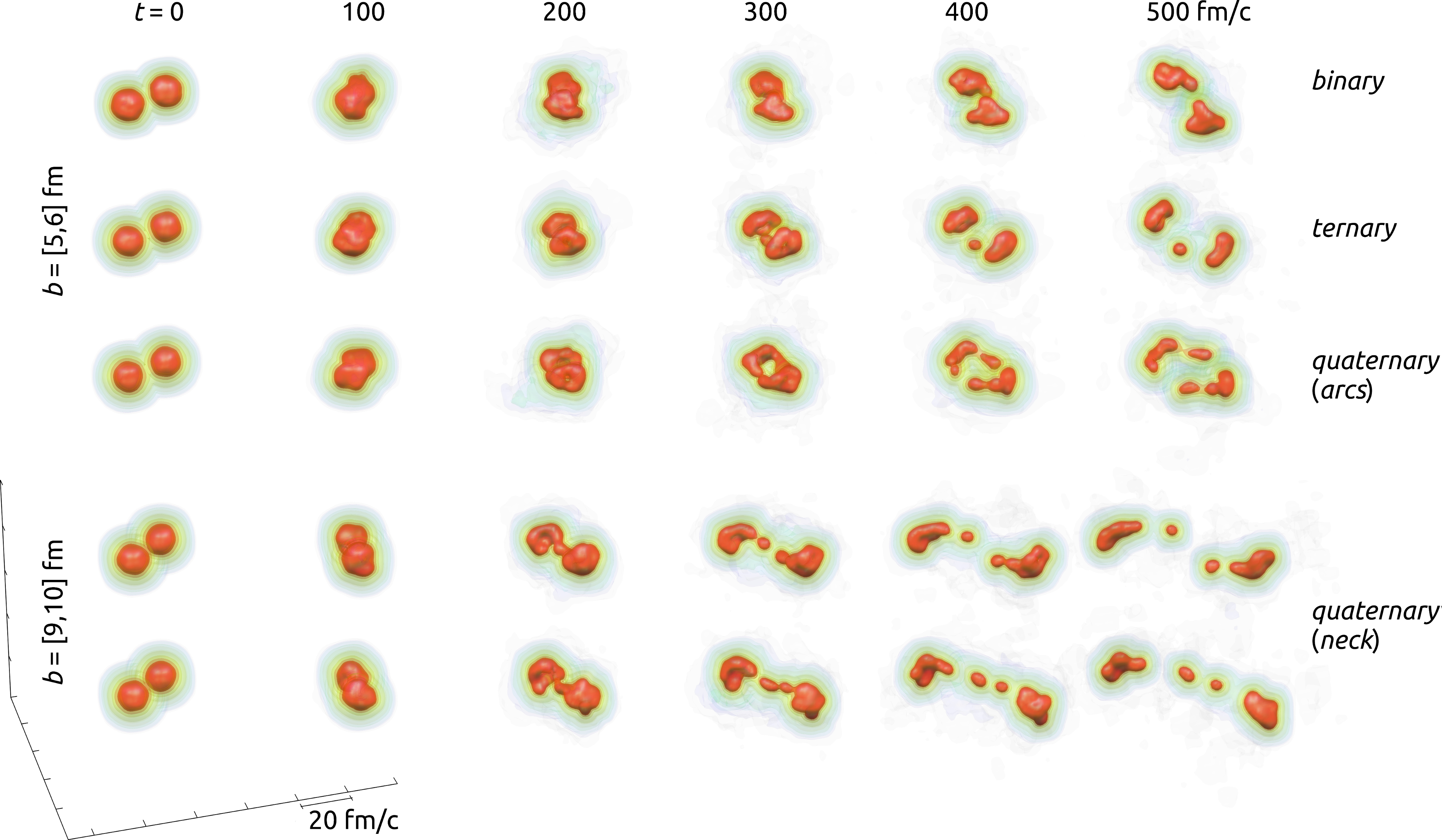}
\end{center}\caption
{
	Three upper rows. Selection of representative events illustrating the three most probable exit channels in the collision $^{197}$Au$+^{197}$Au at 15~$A$~MeV in the interval $b=[5,6]$~fm.
	Most of the configurations at breakup are characterised by two bulges, sometimes connected by one or two arcs developing around a hollow. 
	The separation of the arc thread, in addition to the two bulges imposes ternary and even quaternary fragment configurations.
	Last two rows. Most probable quaternary exit channels in the interval $b=[9,10]$~fm from the breakup of a neck.
}
\label{fig_movie_AuAu15}
\end{figure*}
%
%
\subsection{Configuration-space survey}

	A comprehensive survey on how the BLOB approach describes the system $^{197}$Au$+^{197}$Au at 15~$A$~MeV is shown in Fig.~\ref{fig_fragment_formation_sites_vs_b}.
	Fragment formation sites are registered at $t_{\textrm{stop}}$ for several stochastic events.
	Their distributions in configuration space are projected on the reaction plane and compared for successive intervals of impact parameter $b$ (these fragment-formation-site distributions should not be confused with density distributions in single events, even though they are related).
	The system evolves through a variety of mechanisms as a function of $b$.
	Central collisions undergo some rotation and fragments appear in mostly binary configurations, accumulating in lobes around the PLF and TLF positions.
	In a transition from central to semi-central collisions, fragments start also populating arcs which connect the PLF and TLF lobes.
	In semi-central collisions, fragment formation sites are mainly distributed around a hollow, avoiding the PLF-TLF axis.
	In a transition towards less central collisions, roughly for $b\gtrsim7$~fm, the two arcs gradually merge, until they join into a neck configuration already visible around $b\sim8$~fm, which persists till about $b\sim10$~fm.

\subsection{Fragment multiplicity}

	Fig.~\ref{fig_M_b_stat_survey} presents a study of the fragment multiplicity as a function of the impact parameter $b$.
	The top panel investigates the fragment multiplicity in the early stage that follows the collision, calculated with BLOB up to the time $t_{\textrm{stop}}$.
	In semi-central collisions, direct ternary contributions dominate the cross section and direct quaternary contributions reach their largest rate.
	With increasing impact parameters, firstly quaternary and successively ternary contributions fade in favour of binary channels.

	We focus on the interval $b=[5,6]$~fm because the corresponding configurations at $t_{\textrm{stop}}$ seem to maximise the competition between direct ternary and quaternary splits.
	For this selection, ternary events are the most frequent and they are about three times more frequent than quaternary events (see Fig.~\ref{fig_M_b_stat_survey}, bottom left).
	In the cold system, secondary decays reduce the rate of binary events with the consequence that ternary and quaternary events become the most probable exit channels.
	To complete the survey, we also study the interval $b=[9,10]$~fm, where the quaternary mechanism tends to fade as a direct mode while it is fed by secondary decays.

\subsection{Exotic topologies}	

	The three upper rows of Fig.~\ref{fig_movie_AuAu15} select three representative events illustrating the most probable outcome for the interval $b=[5,6]$~fm.
	Independently of the number of fragments in the exit channel, these configurations tend to drive matter sideward, so that a disk forms along the reaction plane and, eventually, matter distributes around a hollow, while two denser bulges develop around the PLF and TLF positions.
	The configuration leading to four fragments resembles a largely diffused annular thread. 
	For this semi-central selection, the configuration leading to a ternary breakup is not fundamentally dissimilar because the light fragment appearing at midrapidity is also displaced from the PLF-TLF axis.
	This configuration also recalls the study of Ref.~\cite{Rizzo2014} for SMF simulations of a more excited system (23 $A$~MeV); such comparison can make sense because, due to the simplified description of fluctuations, SMF tends to recover at larger energy the results given with BLOB at lower energy, as already observed in the Fermi-energy domain~\cite{Napolitani2013}.
	From this study we infer that in this system a direct breakup into four fragments in a semi-central collision implies the formation of two arcs connecting the PLF and the TLF and not one single neck that disintegrates into two light fragments aligned along the PLF-TLF axis.
	Seemingly, at least for this semi-central impact parameter, the transition from quaternary to ternary breakup corresponds to removing one of the two arcs, so that also in this case the light fragment is misaligned relatively to the PLF-TLF axis.
	As already evident in Fig.~\ref{fig_fragment_formation_sites_vs_b}, 
for more central impact parameters these arcs disappear leaving only the PLF and TLF lobes in a globally binary configuration.

	For semi-peripheral impact parameters the hollow turns into one single thread aligned along the PLF-TLF axis.
	In particular, in the interval $b=[9,10]$~fm few rare events, illustrated in the two bottom rows of Fig.~\ref{fig_movie_AuAu15} may display quaternary patterns along the PLF-TLF axis in a mechanism assimilated to the breakup of a neck.
	In configurations where the neck fragments arise in proximity of the PLF and the TLF some orbiting may bring them slightly off the alignment (last row in Fig.~\ref{fig_movie_AuAu15}).
	The possibility that light fragments could orbit around their close heavy companion was already evoked in Ref.~\cite{Rizzo2014} for semi-central impact parameters. 
	Especially in the deep-inelastic regime, where large deformations are explored, it is therefore necessary to take into consideration the effect of the initial misalignment as a function of the impact parameter (for instance when fragments are formed along arcs), if this mechanism is used as a clock to measure times.
	In the Fermi-energy domain, where timing-versus-kinematics correlations are largely investigated~\cite{Jedele2017,Manso2017}, deformations are less exotic, leading to a reduction of such effect of initial-misalignment as a function of centrality.
	Still, it might be investigated whether other conditions like the larger expansion of the system might produce a similar effect.

\subsection{Kinematics and secondary decay}

%
%
\begin{figure}[b!]
\begin{center}
    \includegraphics[angle=0, width=1\columnwidth]{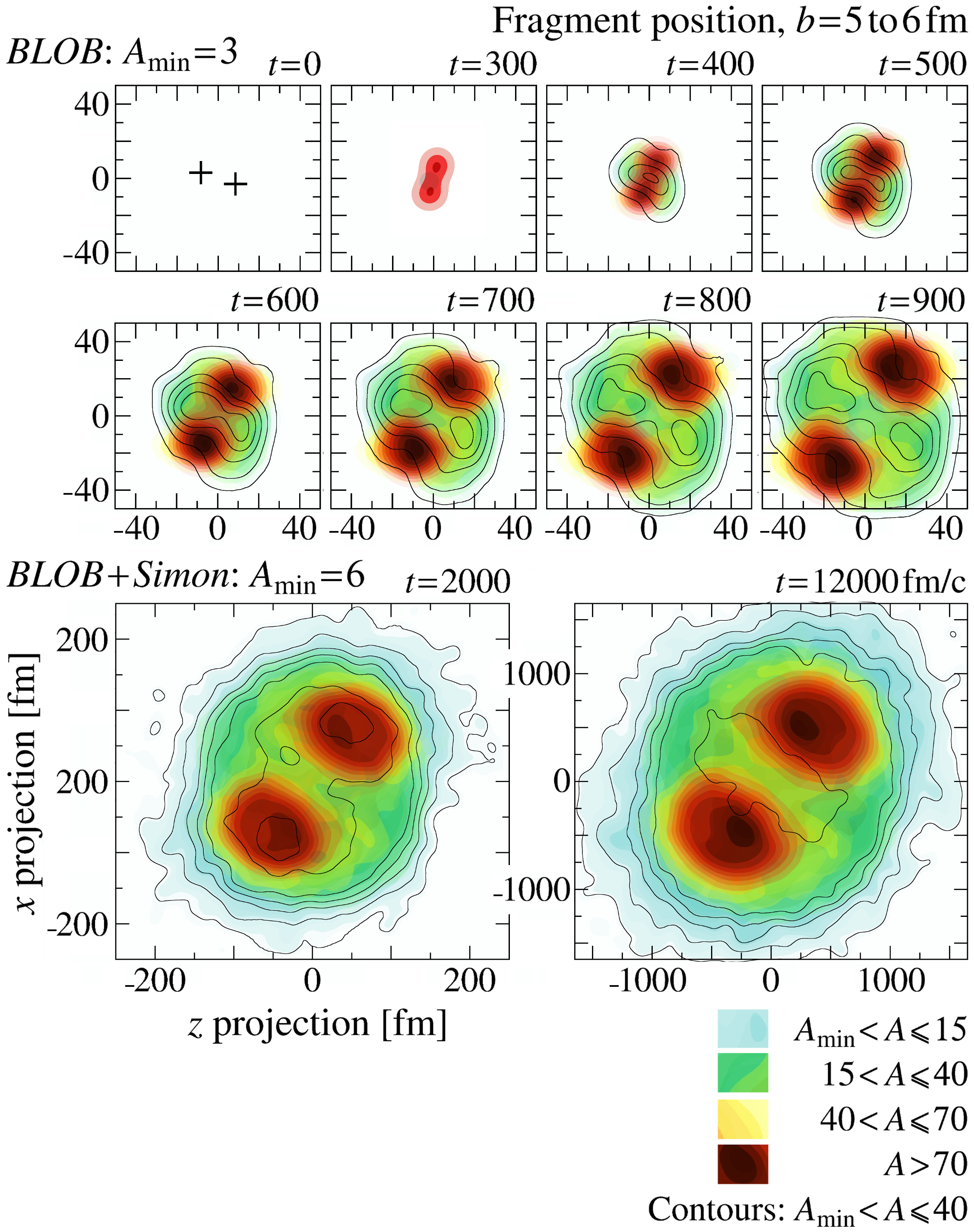}
\end{center}\caption
{
	Projection on the reaction plane of the distribution of fragment formation sites (like in Fig.~\ref{fig_fragment_formation_sites_vs_b}) at successive times, for events corresponding to $b=[5,6]$~fm.
	Colours represent fragment masses; contours select light fragments with $3<A\le 40$ up to $t=900$~fm/c with $6<A\le 40$ for later times.
}
\label{fig_fragment_position_vs_time}
\end{figure}

%
%
\begin{figure}[t!]
\begin{center}
    \includegraphics[angle=0, width=1\columnwidth]{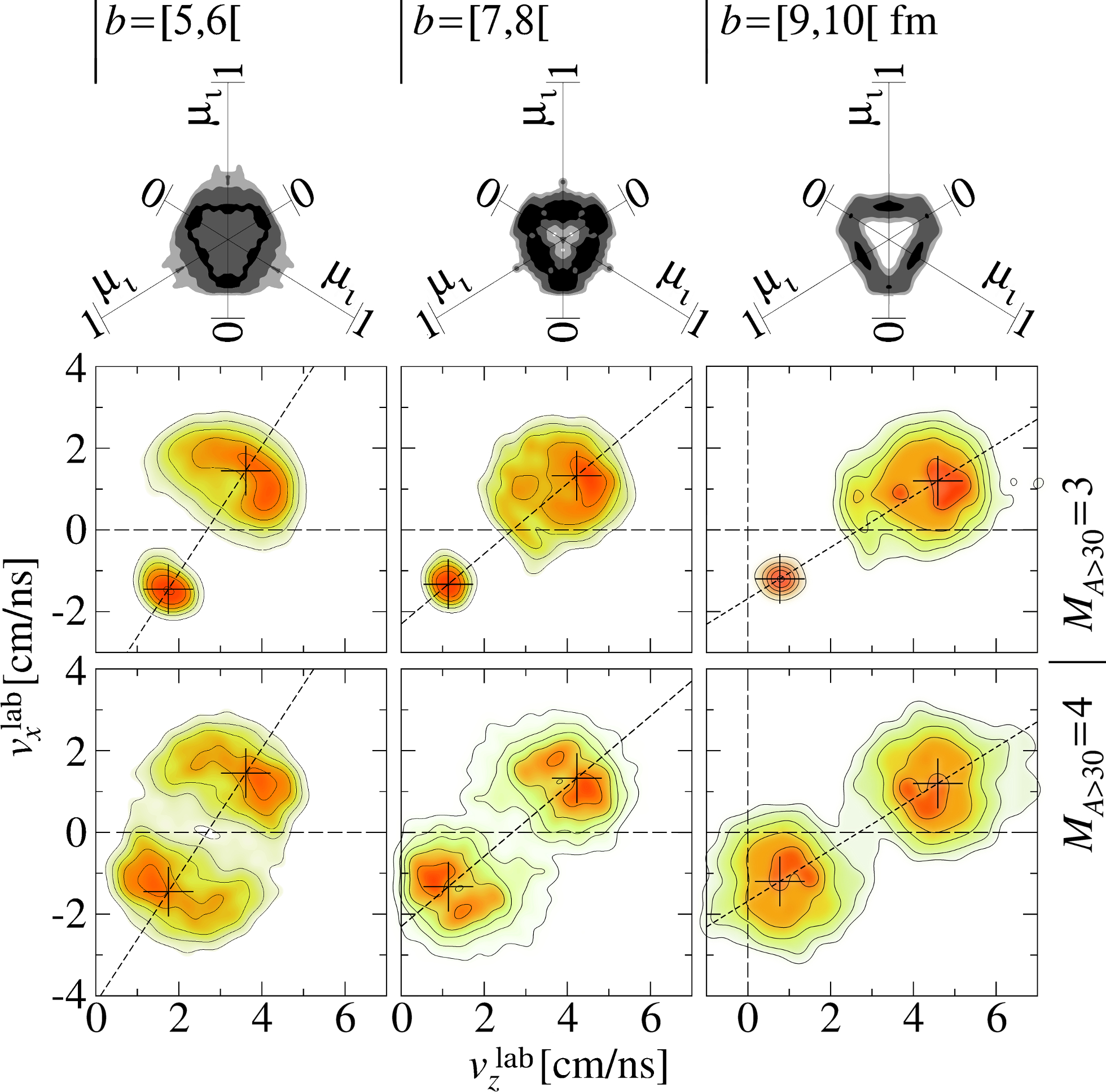}
\end{center}\caption
{
	Velocity distributions projected along the reaction plane ($v^{\textrm{lab}}_z$ along the beam axis, and $v^{\textrm{lab}}_x$ along the impact-parameter axis) for $b=[5,6[$ (left row), $b=[7,8[$~fm (central row) and $b=[9,10[$~fm (right row), calculated after secondary decay in the laboratory frame.
	Splits in three fragments with $A>30$ (middle panels) are presented by assigning negative velocity components to $A_1$ with respect to the centre-of-mass velocity; corresponding Dalitz plots for $\mu_i = A_i/(A_1+A_2+A_3)$ are shown in the upper panels.
	Splits in four fragments with $A>30$ (bottom panels) are presented with symmetrised distributions. 
	Crosses indicate PLF and TLF velocities, calculated in ternary configurations.
}
\label{fig_cold_kinematics_Dalitz}
\end{figure}

	Fig.~\ref{fig_fragment_position_vs_time} investigates the distribution of fragment formation sites projected on the reaction plane (like in Fig.~\ref{fig_fragment_formation_sites_vs_b}), selecting the interval $b=[5,6]$~fm and analysing successive times over the whole reaction process.
	An information on the fragment mass is indicated by the colours and the contours.
	The two upper rows illustrate the evolution of the fragment sites as described within BLOB.
	During these early stages, light fragments ($3<A\le 40$, indicated by contours) appear at midrapidity, signing direct ternary and quaternary events.
	At later times (lower row), when the process is carried on by adding the secondary-decay contribution, especially around 2000~fm/c, light fragments ($6<A\le 40$, indicated by contours) get an additional large contribution around the PLF and TLF lobes from isotropic processes of fission and evaporation.

	According to the present calculation, Fig.~\ref{fig_cold_kinematics_Dalitz} shows velocity distributions and Dalitz plots which could be extracted in an experiment if it where possible to select the impact parameter.
	It exploits correlations among fragments which we label $A_1, A_2, A_3, A_4 \dots$, in decreasing order of size.
	The middle and bottom panels show velocity distributions in the laboratory frame calculated for the cold system and projected on the reaction plane (defined by the beam axis and the impact-parameter axis) for the intervals $b=[5,6]$~fm, $b=[7,8]$~fm and $b=[9,10]$~fm and for ternary and quaternary splits (in case of fragments with $A>30$).
	Ternary splits are presented by assigning negative velocity components with respect to the centre-of-mass velocity to $A_1$, while velocity distributions for quaternary splits are symmetrised with respect to the centre-of-mass velocity; PLF and TLF velocities are indicated by crosses.
	For the ternary splits, corresponding Dalitz plots are also shown.
	The effect of secondary decay is to produce Coulomb rings, visible in the positive-velocity sectors for ternary splits and around both PLF and TLF for quaternary splits.
	Due to the contribution of direct splits, especially for more central impact parameters, the Coulomb rings are distorted and exhibit tails extending towards midrapidity.
	More peripheral impact parameters are associated to larger transparency as evident in the transition from  $b=[5,6]$~fm (corresponding to a hollow configuration like in the example of Fig.~\ref{fig_movie_AuAu15}, second and third row) to $b=[9,10]$~fm (corresponding to aligned configurations like in the last two rows of Fig.~\ref{fig_movie_AuAu15}).
	Especially for ternary splits, larger impact parameters correspond to a loss of isotropy in favour of more aligned configurations.
	Dalitz plots show that in ternary configurations either the PLF or the TLF splits quite independently of the companion; such exit channel becomes less frequent for more central impact parameters, signing a larger contribution from direct splits.
	It may be tempting to compare the study of Fig.~\ref{fig_cold_kinematics_Dalitz} with the experimental results presented in Refs.~\cite{SkwiraChalot2008,Wilczynski2010b,Wilczynski2010a}, after some caution is kept as no experimental-device filter is applied to the study of Fig.~\ref{fig_cold_kinematics_Dalitz}. 
	We can point out that both the calculation and the experimental data agree in showing non-isotropic contributions for ternary and quaternary splits and that those latter present very similar features especially for the range of larger impact parameters (which contribute the most to the geometric cross section).
	The filling of midrapidity is due to the anisotropy of dynamical contributions which can be explicitly treated in the present model and is more important for more central impact parameters.
	Also Dalitz diagrams, especially for $b=[9,10]$~fm are similar to the experimental results.

\subsection{Fragment production}
%
%
\begin{figure}[t!]
\begin{center}
    \includegraphics[angle=0, width=.95\columnwidth]{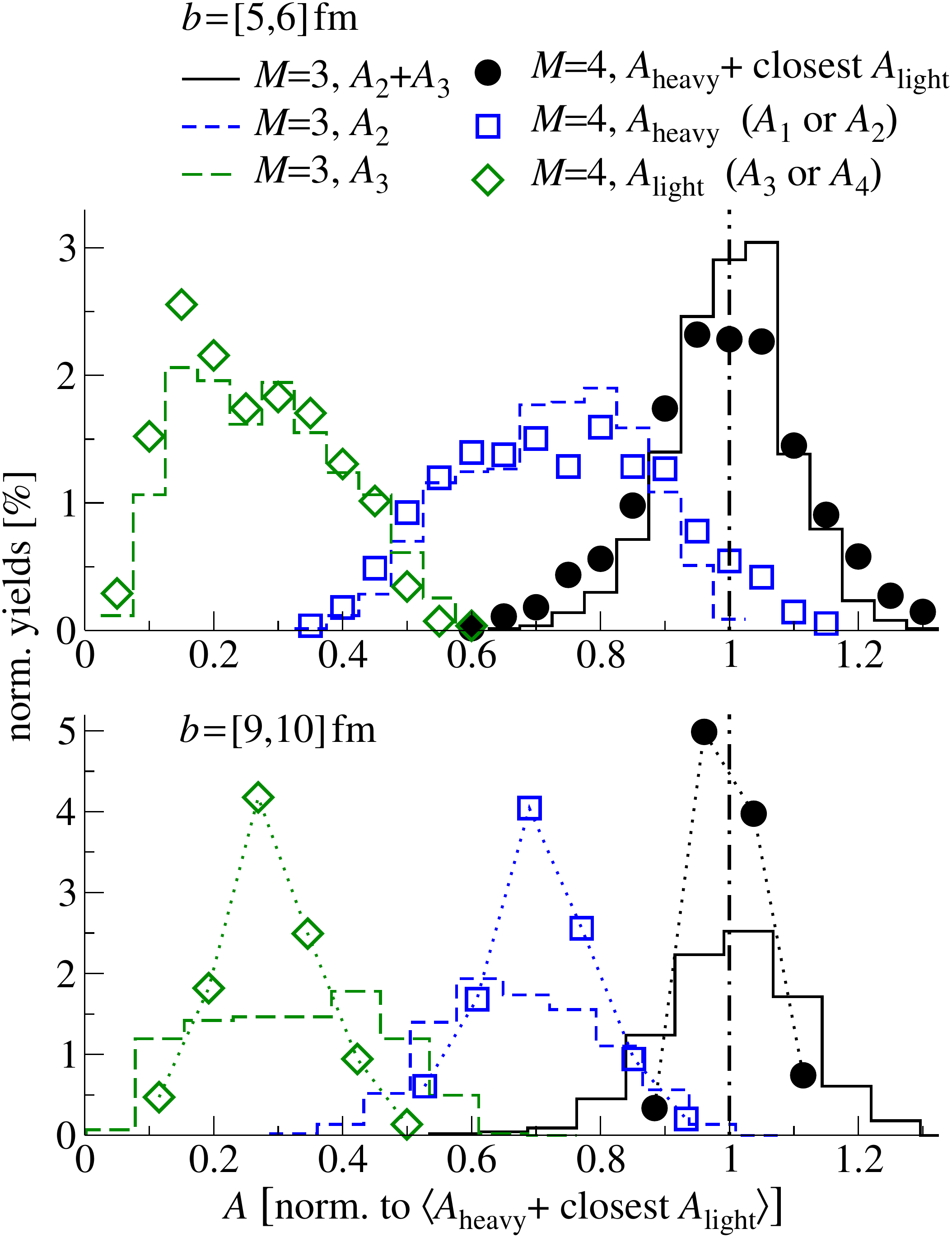}
\end{center}\caption
{
    Mass distributions calculated with BLOB for ternary ($M3$, dashed histograms) and quaternary events ($M4$, symbols) corresponding to $b=[5,6]$ (top) and $b=[9,10]$~fm (bottom). 
	Green histograms represent the distribution of light fragments, blue histograms represent the distribution of the corresponding heavy partners, and black histograms are the sum of the heavy- and light-partner contributions.
	Masses are normalised to the mean value of this latter (black) histogram.
	Yields are normalised to the total production yield restricted to the given impact-parameter contribution and selected fragment multiplicity.
	Ternary and quaternary mechanisms result closely comparable, see text.
}
\label{fig_A1_A2_A3_A4_distributions}
\end{figure}
%
%
%
\begin{figure}[b!]
\begin{center}
    \includegraphics[angle=0, width=1\columnwidth]
{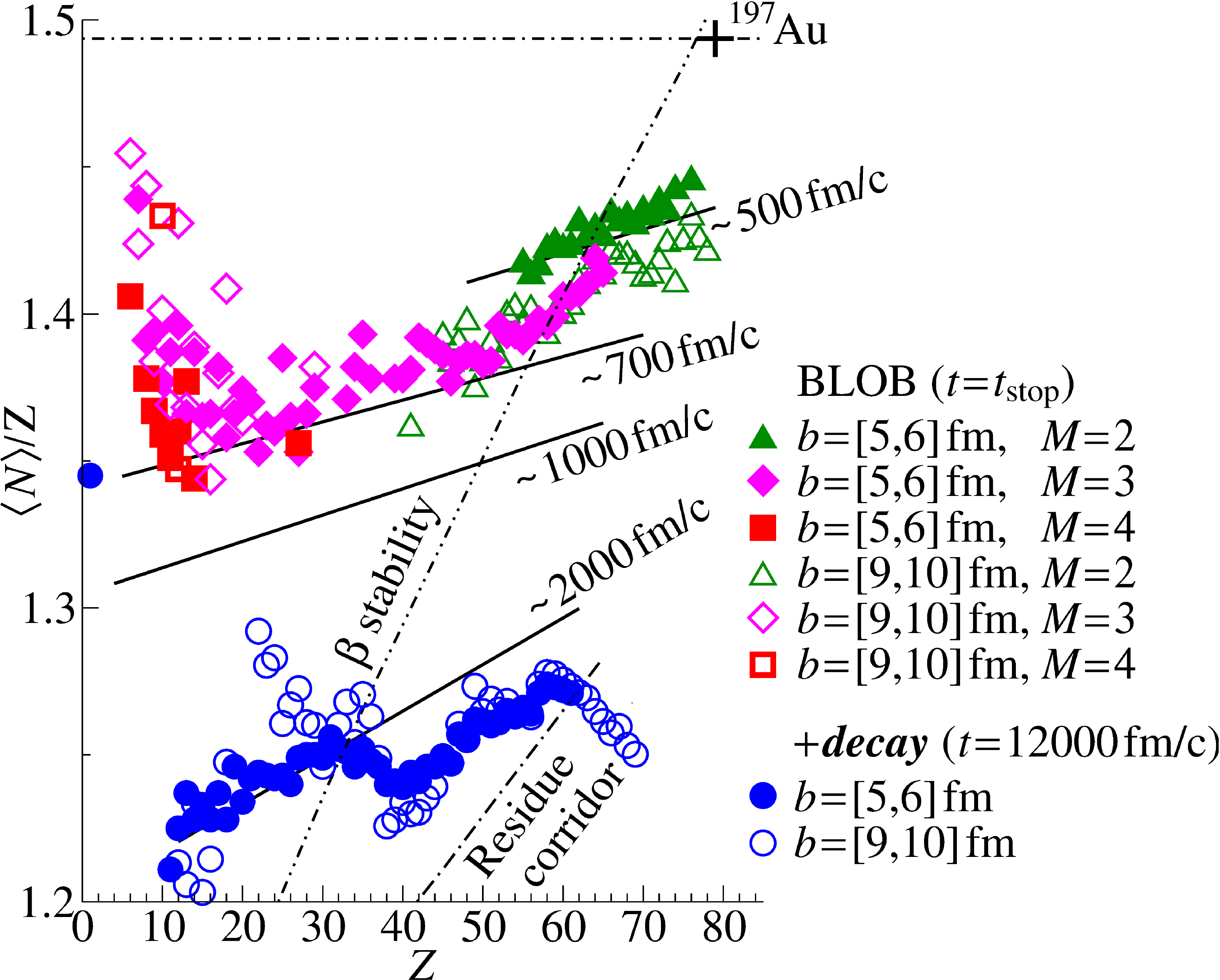}
\end{center}\caption
{
    Mean neutron-to-proton ratio $\langle N\rangle/Z$ of fragments as a function of $Z$ at successive times for events corresponding to $b=[5,6]$~fm (filled symbols) and $b=[9,10]$~fm (open symbols).
	The hot production (simulated with BLOB) is separated into multiplicity contributions (binary ternary and quaternary for green violet and red symbols, respectively).
	The final production, obtained after secondary decay (blue symbols) is extracted at $t=12000$ fm/$c$.
	Solid lines are fits giving indicative information on time evolution.
    }
\label{fig_NoverZ_vs_Z_time}
\end{figure}
%
%

%
%
	Focusing on the direct ternary-quaternary competition in semi-central collisions, Fig.~\ref{fig_A1_A2_A3_A4_distributions} presents a study of the light-fragment sizes in the spirit of the study proposed in Ref.~\cite{Rizzo2014}.
	Selecting the intervals $b=[5,6]$ (top) and $b=[9,10]$~fm (bottom), normalised mass distributions are calculated with BLOB at time $t_{\textrm{stop}}$ for ternary and quaternary events. 
	Splits occur along arc threads with heavier residues close to the TLF/PLF position and lighter companions at midrapidity.
	For each ternary or quaternary event the green distribution accounts for one (for ternary events) or two (for quaternary events) light fragments, the blue distribution accounts for the corresponding closest heavier partners in configuration space, and the black distribution accounts for couples of light and heavy partners by collecting their total mass.
	All distributions are then normalised to the mean value of the summed light and heavy fragment sizes (black distribution). 
	Yields are normalised to the fraction of total production obtained for the given impact-parameter contribution and selected fragment multiplicity.
	This result is quantitatively closely comparable to the study of Ref.~\cite{Rizzo2014}, with the difference that the present analysis profits from completely separated fragments issued from the BLOB treatment.
	The present analysis reveals a similarity between ternary and quaternary mechanisms, by indicating that the properties of a single light fragment are not influenced by the possible presence of a second light companion.
	In a quaternary semi-central mechanism, the two light fragments would in fact form simultaneously and independently along two separate arcs and the process would still be unchanged if only one arc were present in a ternary mechanism.
	Seemingly, in a quaternary semi-peripheral mechanism, the two light fragments would form along the TLF-PLF axis but their multiplicity does not affect the mean values of their distributions.

%
%
	Fig.~\ref{fig_NoverZ_vs_Z_time} investigates for the selection $b=[5,6]$ and $b=[9,10]$~fm the evolution in time of the mean neutron-to-proton ratio $\langle N\rangle/Z$ of fragments as a function of $Z$.
	For the early stages of the process, at $t_{\textrm{stop}}$ (ranging therefore over the full $t_{\textrm{stop}}$ window, see Fig.~\ref{fig_Probability_split_b5to6_b9to10}), 
the fragment production simulated with BLOB is shown divided into binary, ternary and quaternary contributions.
	Even though the production associated to ternary and quaternary contributions is very neutron rich, in general it reflects the isotopic composition $\langle N\rangle/Z$ of the mother nuclei, like in fission.
	The evolution of the isotopic composition up to $t=2000$ fm/$c$ is schematically sketched by linear fits to $\langle N\rangle/Z$ distributions calculated at corresponding times in the $b=[5,6]$~fm window.
	After secondary decay, at $t=12000$ fm/$c$, the final isotopic composition is more neutron rich than the evaporation residue corridor and tends to accumulate around $\beta$ stability.
	From this information alone and without a survey of density gradients in the process, we can not recognise the signature of an isospin drift process. 
	At larger incident energy, and typically in the Fermi-energy regime, the fragmentation of a neck would be related to isospin drifts driven by density gradients and to the neutron enrichment of low-density regions~\cite{Lionti2005,Baran2005,DiToro2006,Colonna2017}.
	This density dependence is investigated in the following.

\section{Identifying instabilities behind dynamical splits \label{discussion}}

	Consistently with experimental results, the Boltzmann-Langevin approach and the analysis of section~\ref{results} describe mechanisms where three and four fragments are formed.
	We give thereafter some interpretation related to transport observables with a focus on the interval $b=[5,6]$~fm, where dynamical contributions are the most prominent, and $b=[9,10]$~fm, where aligned patterns of three or four fragments and the related kinematics match experimental observations.

\subsection{Isospin-density-temperature analysis}

%
%
	The isotopic composition of fragments, analysed in Fig.~\ref{fig_NoverZ_vs_Z_time} and in section~\ref{results}, exhibits a tendency to enhanced neutron enrichment for the lightest fragments produced at early times.
	This likely reflects the isotopic composition of the breaking-up source, like in fission, and it may additionally include possible contributions from isospin drift, as observed in collisions at Fermi energy, if density gradients are involved in the process.

%
%
	To clarify this aspect, we went through a more specific analysis of the density dependence of the clusterisation process along the neck and arc threads and we investigated the type of related instability.
	Fig.~\ref{fig_T_vs_rho_yields_propto_symbolarea} tracks the temperature versus density correlation calculated with BLOB for ternary and quaternary events corresponding to the interval $b=[5,6]$ (left) and $b=[9,10]$~fm (right).
	Colours distinguish masses below or above $A=30$; 
	Temperature and density are extracted from the portions of kinetic-energy and density distributions which correspond to a given fragment at the time of formation along a neck or arc thread (PLF and TLF are excluded).
	Symbol areas are linearly proportional to the yields.
	The line is a fit to the calculation for $b=[5,6]$~fm, for comparison.
	It should be noted that, even though at the same time $t_{\textrm{stop}}$ more central collisions are associated to larger temperatures, the temperature ranges corresponding to the two intervals $b=[5,6]$ and $b=[9,10]$~fm result similar because they are registered at different intervals of time (i.e. earlier $t_{\textrm{stop}}$ for larger $b$).
	The density oscillates slightly around saturation, without dropping to critically low values so that, also from this point of view, volume instabilities could be excluded.
	Semi-peripheral channels present a larger density variance because smaller fragments ($A<30$) experience lower densities.
	Even though these densities are not sufficient to trigger isospin drifts, like at Fermi energies, this slight density drop might sign an initial tendency in such direction.

\subsection{Linear-response analysis of unstable modes}
%
%
\begin{figure}[pt!]
\begin{center}
    \includegraphics[angle=0, width=1.\columnwidth]{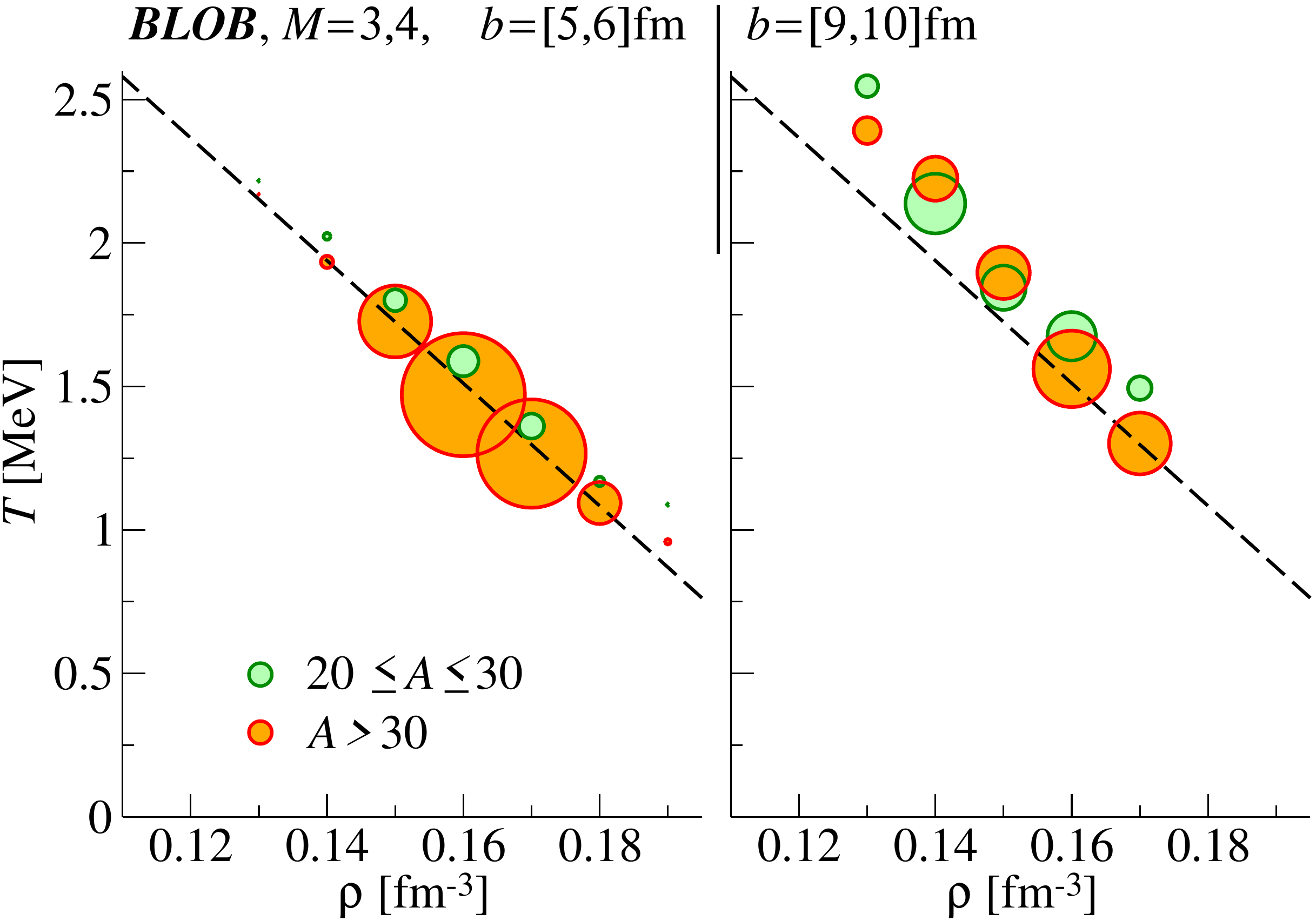}
\end{center}\caption
{
	Temperature-versus-density correlation in ternary and quaternary collisions for $b=[5,6]$~fm (left) and $b=[9,10]$~fm (right) for fragments with $A\ge 20$ at the time of formation, developing along a neck or arc thread.
	Colours distinguish fragment masses below or above $A=30$.
	Symbol areas are linearly proportional to the yields.
	The line is a fit to the $b=[5,6]$~fm configuration.
}
\label{fig_T_vs_rho_yields_propto_symbolarea}
\end{figure}
%

\begin{figure}[b!]
\begin{center}
    \includegraphics[angle=0, width=.8\columnwidth]{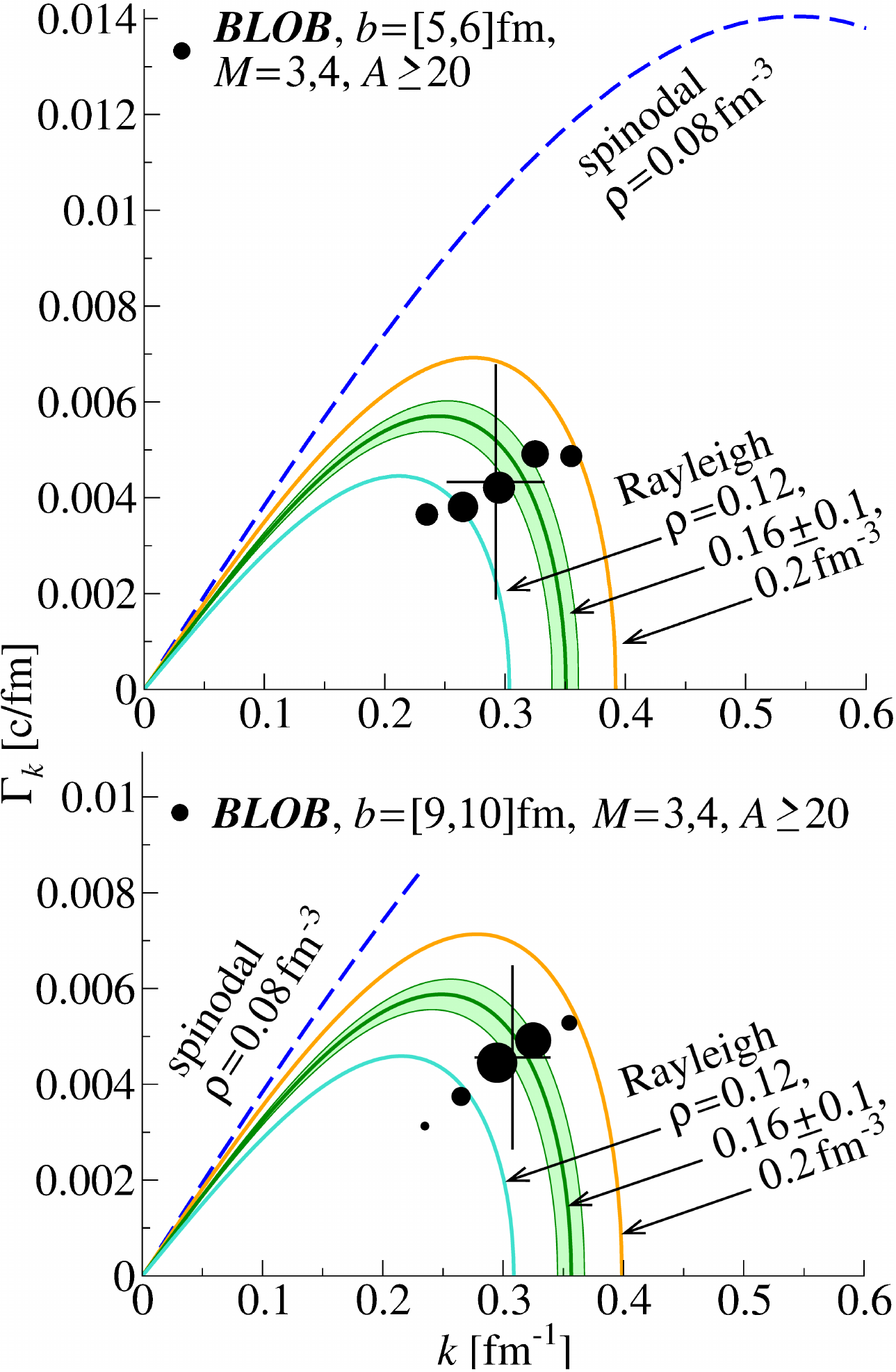}
\end{center}\caption
{
	Growth rate of unstable modes as a function of the corresponding wave number at temperature and densities compatible with the study of Fig.~\ref{fig_T_vs_rho_yields_propto_symbolarea}.
	The dashed lines illustrate the dispersion relation in nuclear matter for the spinodal (volume) instability at half saturation density.
	For different densities, solid lines illustrate the dispersion relation for the Rayleigh (surface) instability compatible with $^{197}$Au$+^{197}$Au at 15~$A$~MeV for $b=[5,6]$~fm (top) and $b=[9,10]$~fm (bottom).
	For the same systems, dots with an area linearly proportional to the count rate indicate the $k$ modes which dominate the instability as obtained from BLOB simulations.
	Crosses are averages of all $k$ contributions with the corresponding standard deviation.  
}
\label{fig_dispersion_relation_Rayleigh_spinodal}
\end{figure}

%
%
	Even more specifically, we carried on a linear-response analysis in the spirit of Ref.~\cite{Napolitani2019}.
	Fig.~\ref{fig_dispersion_relation_Rayleigh_spinodal} presents a study of the growth rate of unstable modes as a function of the corresponding wave number $k$.
	The analytic dispersion relation calculated for the Plateau-Rayleigh (surface) instability is obtained from the following analytic prescription~\cite{Brosa1990}:
\begin{equation}
	(\Gamma_{k, \textrm{surf}})^2 = 
	\frac{\gamma(\rho,\beta,T)}{\rho m r^3} \frac{I_1(kr)}{I_0(kr)} kr (1-k^2r^2)
	\;,
	\label{eq:Rayleigh}
\end{equation}
which describes the growth rate $\Gamma_{k, \textrm{surf}}$ for surface fluctuations of wave number $k$ developing along a columnar thread of radius $r$ at a given local density $\rho$.
	$r$ is the radius of the arc or neck thread linking a light fragment ($A_3$ or $A_4$) to its closest heavy companion ($A_1$ or $A_2$).
	$I_0$ and $I_1$ are modified Bessel functions and $m$ is the nucleon mass.
	Eq.~(\ref{eq:Rayleigh}) is slightly modified with respect to the original prescription of Ref.~\cite{Brosa1990} by setting the surface tension $\gamma(\rho,\beta,T)$ dependent on density, charge symmetry $\beta=(\rho_n-\rho_p)/\rho$ (using the prescription suggested in refs.~\cite{Iida2004,Horiuchi2017} for the SKM$^{*}$ interaction), and temperature $T$ (introducing the correction $F_T$ proposed in Ref.~\cite{Ravenhall1983}):
\begin{equation}
	\gamma(\rho,\beta,T)
	\approx F_T \Big[
	1 -c_{\textrm{sym}}\beta^2 
		-\chi\Big(1-\frac{\rho}{\rho_{\textrm{sat}}}\Big)	\Big]
		\gamma_{\textrm{sat}} 
	\;.
	\label{eq:surftension}
\end{equation}
The effect of a finite temperature below $3$~MeV is negligible so that we can take $F_T\!\approx\!1$.
According to the SKM$^{*}$ interaction, 
$c_{\textrm{sym}}\!\approx\!1.9$,
$\chi=(\rho_{\textrm{sat}}/\gamma_{\textrm{sat}})\partial_\rho \gamma|_{\rho_n=\rho_p=\rho_{\textrm{sat}}/2}\approx1.16$.
$\rho_{\textrm{sat}}\approx 0.16$~$\text{fm}^{-3}$, and
$\gamma_{\textrm{sat}}=\gamma(\rho_{\textrm{sat}},\beta\!=\!0,T\!=\!0)\approx 1 $~$\text{MeV\,fm}^{-2}$.
	Other effects, like diffuseness and viscosity~\cite{Baldo2012} (which have counteracting contributions~\cite{Brosa1990}) as well as geometric distortion, are neglected.
%
%
	Eq.~(\ref{eq:Rayleigh}) is plotted in Fig.~\ref{fig_dispersion_relation_Rayleigh_spinodal} (green solid lines) for both semi-central (top) and semi-peripheral (bottom) collisions. 
	For $b=[5,6]$fm we used a value of $r$ ($2.85\pm0.56$ fm) averaged on all calculated stochastic events, at a temperature $T$ ($1.47\pm0.21$ MeV) and local density $\rho$ ($0.16\pm0.01$ fm$^{-3}$) measured as the density averaged over the potential well associated to the light fragment at the instant of separation; the charge asymmetry $\beta$ corresponds to the maximum of the breakup probability ($0.174\pm0.006$).
	These parameters are obtained from a numerical shape analysis, where neutron and proton density distributions are tracked as a function of time for each stochastic event.
	For neck threads formed at $b=[9,10]$fm we used very similar values 
($r=2.9\pm0.6$ fm, $T=1.78\pm0.44$ MeV, $\rho=0.15\pm0.01$ fm$^{-3}$, and same $\beta$).
	Density and temperature are compatible with the study of Fig.~\ref{fig_T_vs_rho_yields_propto_symbolarea}.
	The standard deviations of all the above quantities contribute to the calculation uncertainty, represented by the filled band around the Rayleigh dispersion relation for $\rho=0.16$ fm$^{-3}$.
	Such uncertainty is completely determined by density, while temperature and charge asymmetry yield irrelevant contributions.
	Two additional calculations trace the Rayleigh dispersion relation for two extreme values of the density (not attained), $\rho=0.12$ and $0.2$ fm$^{-3}$.
%
%
%
%

%
%
	For comparison, the dashed line illustrates the analytic dispersion relation calculated in nuclear matter for the spinodal (volume) instability at half saturation density.  
	We may note that it could not be established for larger densities because the conditions for the spinodal instability are no more matched.
	The same interaction properties determine diffuseness and surface tension in the Plateau-Rayleigh modes, as well as low-density volume-instabilities in the spinodal modes.
%
%
	The analytic dispersion relations are compared to growth rates extracted from the numerical simulation of $^{197}$Au$+^{197}$Au at 15~$A$~MeV for events corresponding to the interval $b=[5,6]$ (top) and $b=[9,10]$~fm (bottom) and leading to ternary or quaternary splits with fragments of $A\ge 20$.
	A Fourier analysis of density fluctuations cannot be applied to the open system under consideration.
	Hence, we extracted the growth rate directly from the action of fluctuations on cluster correlations and the associated chronology. 
	We estimated the rupture time of the arc threads, $t_{\textrm{rupture}}$, as the average time when light fragments separate from the thread, calculated since inhomogeneities start to arise (i.e. around $\sim 350$ and $\sim 240$ fm/c for the interval $b=[5,6]$ and $b=[9,10]$~fm, respectively) 
	Consequently, the growth rate is evaluated as $\Gamma_{k} = \hbar / t_{\textrm{rupture}}$.
	The wavelength of unstable modes is assumed to correspond to the average distance between a light fragment ($A_3$ or $A_4$) and the corresponding heavy partner ($A_1$ and $A_2$), measured along the arc or neck thread.
	Such approach, firstly introduced in Ref.~\cite{Napolitani2017} for spinodal fragmentation and then used in Ref.~\cite{Napolitani2019}, estimates $\Gamma_{k}$ satisfactorily in realistic physical conditions, at the price of some possible underestimation.
	From this picture, circles with an area linearly proportional to the count rate indicate the $k$ modes which dominate the instability growth.
	The crosses indicate the average of all wave number contributions with the corresponding standard deviations.
%
%
	From the above procedure, we found that the instability which triggers clusterisation along the neck or arc threads has a growth rate and a wavelength range closely compatible with the Rayleigh instability for density conditions in the proximity of saturation.
	At the same time, this instability is largely incompatible with a volume (spinodal) instability; this latter would be faster, it would involve smaller wavelengths (and produce smaller fragments) and act at about three times lower density.
	The Rayleigh surface instability is usually associated to fission~\cite{Brosa1990}.
	In the case of these ternary and quaternary mechanisms, however, the process is faster than ordinary fission and it has a clear kinematic signature, depending on the exotic fragment configuration, and is reflected in corresponding angular correlations.

\section{Conclusions}

	We applied the Boltzmann-Langevin One Body approach to describe shape fluctuations and exotic fragmentation modes in very heavy nonfusing systems in the deep-inelastic regime.
	At variance with the Fermi-energy domain, where multiple-fragment modes are frequent, in the deep-inelastic regime instabilities are not associated with expansion effects but they rather characterise highly deformed and elongated configurations.
	In such conditions, the BLOB approach of Eq.~(\ref{eq:BLOB}) provides a thorough description of shape fluctuations leading to fragment formation.
	As a case study, we took the system $^{197}$Au$+^{197}$Au at 15~$A$~MeV, which has been the subject of dedicated experiments~\cite{SkwiraChalot2008,Wilczynski2010b,Wilczynski2010a}, and undertook a simulation of transport observables in connection with fragment formation and instability propagation.

	We found that, in semi-central collisions, where fusion channels are suppressed, the system explores exotic noncompact shapes associated to the formation of a sort of annular thread flattened along the reaction plane.
	From this configuration at least two bulges form along the annular thread in correspondence with the PLF/TLF positions, leading to binary splits.
	Rather frequently, one or two more fragments separate from the arcs which connect the PLF/TLF bulges, leading to direct ternary and quaternary splits occurring in comparable conditions and, by construction, misaligned with respect to the PLF-TLF axis.
	From this process, in semi-central collisions an interesting competition between direct ternary and quaternary splits dominates the cross section.
	With increasing impact parameter, in a transition from semi-central ($b=[5,6]$~fm) to semi-peripheral ($b=[9,10]$~fm) collisions we observe a gradual stretching of the annular thread along the PLF-TLF axis until the system develops a long neck connecting the PLF and TLF bulges.
	Along this transition towards semi-peripheral collisions, direct multiple breakup channels progressively fade, and direct ternary channels become more frequent than direct quaternary channels.
	Still, direct quaternary splits can be observed till roughly $b=10$~fm as emerging from the same neck-like thread aligned along the PLF-TLF axis.
	Larger multiplicities are however recovered from secondary fission-like processes which in our present study have been described through a statistical afterburner (SIMON) which also accounts for the Coulomb field produced by the ensemble of particles and fragments composing the system in flight.
	As discussed in section~\ref{collision}, to achieve an even improved description of kinematic features (like collinearity) when secondary stages are involved, a future challenge should be the dynamical description of the whole process, where deformations of heavy PLF and TLF could be taken into account also at late times.

	From the analysis of unstable modes, we found that both the annular thread and the neck thread rupture in comparable conditions, at a density which slightly oscillates around saturation density and due to a surface instability of Rayleigh type which is incompatible with the volume (spinodal) instability which characterise Fermi energies.
	In comparison with nuclear spinodal instability, from a dispersion-relation study of the corresponding unstable modes $k$ (see Fig.~\ref{fig_dispersion_relation_Rayleigh_spinodal}, in comparison with the study of Ref.~\cite{Napolitani2017}), we concluded that for both semi-central and semi-peripheral impact parameters the leading wavelength related to the surface instability is roughly twice larger ($\lambda=20~fm$) and the leading growth time can be up to ten times longer ($t_{\textrm{rupture}}=\hbar/\Gamma_{k}\approx$ 200 to 300~fm/c).
	As a consequence, rather massive fragments (roughly up to Germanium) are produced in such process with a formation time which, calculated from the onset of instabilities, extends over roughly 500~fm/c.
	Such time window is so large to overlap with the secondary decay process.
	In this respect, due to the large growth time of the leading instability and the large spanning of the fragment formation time, the transition from direct to secondary processes does not present any clear separation into two distinct stages (see Fig.~\ref{fig_Probability_split_b5to6_b9to10}).
	Rather, both direct and secondary processes involve a prominent dynamical character related to the elongated configuration of the system.
	This analysis of shape fluctuations and instabilities explains the kinematic anisotropies and the fast dynamics which have been reported in experiments.

\section{Acknowledgements}

Research was conducted under the auspices of the International Associated Laboratory (LIA) COLL-AGAIN.
	Funding from the European Union’s Horizon 2020 research and innovation program under Grant No. 654002 is acknowledged.

\end{document}